\documentclass[prb,showpacs,superscriptaddress,twocolumn,amssymb,amsfonts]{revtex4-1}
\usepackage{amsmath}
\usepackage{bm}
\usepackage{epsfig}
\usepackage{graphics}
\usepackage{dsfont}

\def\beq{\begin{equation}}
\def\eeq{\end{equation}}
\def\beqn{\begin{eqnarray}}
\def\eeqn{\end{eqnarray}}

\renewcommand{\bf}{\mathbf}

\begin{document}
\title{Possible topological phases of bulk magnetically doped Bi2Se3: turning a topological band insulator into Weyl semimetal}
\author{Gil Young Cho}
\affiliation{Department of Physics, University of California,
Berkeley, CA 94720}
\date{\today}


%

\begin{abstract}
We discuss the possibility of realizing Weyl semimetal phase in the magnetically doped topological band insulators. When the magnetic moments are ferromagnetically polarized, we show that there are three phases in the system upon the competition between topological mass and magnetic mass: topological band insulator phase, Weyl semimetal phase, and trivial phase. We explicitly derive the low energy theory of Weyl points from the general continuum Hamiltonian of topological insulators near the Dirac point, e.g. ${\bf k} \cdot {\bf p}$ theory near $\Gamma$ point for Bi$_{2}$Se$_{3}$. Furthermore, we introduce the microscopic tight-binding model on the diamond lattice to describe the magnetically doped topological insulator, and we found the Weyl semimetal phase. We also discuss the dimensional cross-over of the Weyl semimetal phase to the anomalous Hall effect. In closing, we discuss the experimental situation for the Weyl semimetal phase.
\end{abstract}


\maketitle
The discovery of topological band insulators~\cite{hm, mb, fkm, roy} has been bringing attention to the topological properties of the band insulators. These phases are insulating in the bulk but support metallic surface states which is protected as far as the bulk gap remains open and time-reversal symmtry is respected. The topological properties of topological band insulators~\cite{qi1, fkm, axion_Essin} are guaranteed due to the well-defined bulk gap. However, it's demonstrated recently that gapless Weyl semimetal phase can also have topological protection to opening up a gap in bulk, edge-bulk correspondence, and gapless anomalous Hall response~\cite{Ashvin, Yang, Burkov, Fang}.

In this paper, we present that topological insulator materials can be turned into the Weyl semimetal phase upon doping the bulk with magnetic materials that is assumed to be ferromangetically ordered. Then, the magnetization mass and topological band gap start to compete and result three phases including topological Weyl semimetal phase. The Weyl semimetal phase shows up when the magnetization mass is  strong enough to close the topological band gap. This is striking in that in the previous studies on the effect magnetization (where magnetization mass is usually taken as the small `perturbation'), it's thought not desirable to close the band gap as it will destroy the topological properties of the topological band insulator and turn the material into trivial one. We hereby show that the strong magnetization mass in three-dimensional topological band insulator materials can generate a new topological phase; Weyl semimetal phase. The other two phases are topological band insulator phase (characterized by the axion angle $\theta = \pi$) and trivial insulator phase ($\theta =0$). We show that the gapless Weyl phase is relatively easy to realize in the topological band insulator in experiment (See also the multilayered structure with topological insulators~\cite{Burkov, Burkov2} which uses the surface bands of topological insulators and recently proposed material HgCr$_{2}$Se$_{4}$~\cite{Fang}), compared to the complex compound structure A$_{2}$Ir$_{2}$O$_{7}$~\cite{Ashvin, Yang, Yong} where the possibility of realizing Weyl semimetal phase in condensed matter system is proposed. Our continuum theory for the Weyl semimetal phase is directly applicable to the well-known topological insulator Bi$_{2}$Se$_{3}$~\cite{BiSe}. Our work would be perhaps most relevant to the material TlBi(S$_{1-x}$Se$_{x}$)$_{2}$ where the gap could be tuned to the criticality in experiment~\cite{BiSeS}. This continuum theory for the general low energy theory of the topological insulators is supported for the standard tight-binding model on the diamond lattice. Our discussion is then extended to the dimensional cross-over of the Weyl semimetal phase to the two-dimensional anomalous Hall effect.  

\section{Theoretical Perspective}
The universal low-energy physics of the topological insulator materials can be captured in the massive (isotropic) Dirac system
\beq
H = v \tau^{z} \sum_{a=1}^{3}\sigma_{a} {\bf k^{a}} + M({\bf k}) \tau^{x}
\label{TI}
\eeq   
where $\tau^{\mu}$ and $\sigma^{\nu}$ are Pauli matrices. We treat $\sigma^{\mu}$ as the `real' spin which is relevant in dealing with time-reversal symmetry. The Hamiltonian Eq.\eqref{TI} incorporates with the two discrete symmetries: the time-reversal symmetry and the inversion symmetry. Usually, the Dirac mass $M$ has the momentum dependence, i.e., $M = M_{0} + b {\bf k}^{2}$ which can be tuned to be in the topological band insulator phase if $M_{0}b <0$ and trivial phase if $M_{0}b >0$~\cite{}. When we are in the topological insulator phase~\cite{fkm}, there is a single Dirac cone on the surface
\beq
H_{s} = \pm v \sum_{i=1}^{2}\sigma^{i} \cdot {\bf k}_{i} 
\eeq
which can support a {\it quantized} anomalous Hall effect~\cite{qi1} upon introducing time-reversal symmetry breaking gap $\sim m \sigma^{z}$ on the surface. This effect attracts enumerous attention~\cite{QAH, QAH1, axion_monopole, Franz_FM} which is related to ``axion electromagnetism''~\cite{qi1, axion_Essin}. The effect can happen when the magnetization gap $\sim m$ is safely smaller than $M_{0}$ so that we are in the topological band insulator phase. 

We further ask what would be the effect of the magnetization mass if it turns large enough (due to the bulk magnetic impurity doping). We assume that the magnetic material doped in the topological band insulator forms a ferromagnetic order to supply the necessary mass. If this magnetization is large enough, then strikingly it can generate a gapless Weyl semimetal phase. To study this, we introduce the magnetization mass in the form of
\beq
H_{m} = J{\bf m} \cdot \sigma \tau^{0}
\eeq  
to the Hamiltonian Eq.\eqref{TI}. Due to the isotropy in our model, we can further simplify $J{\bf m} = m {\hat z}$ (redefining $Jm \rightarrow m$). As a whole, we will deal with the Hamiltonian 
\beq
H = v \tau^{z} \sum_{a}\sigma_{a}{\bf k^{a}} + M({\bf k}) \tau^{x} + m\sigma^{z}\tau^{0}
\label{Weyl}
\eeq
which supports three phases: topological band insulator with time-reversal symmetry broken surface, Weyl semimetal phase, and trivial phase (without losing generality, we choose $M_{0}>0$ and $m>0$). We will show that the transition between two insulating phases and gapless Weyl semimetal phase happens at $M_{0}=m$.   

Upon with this general consideration, we start to discuss the details of the Hamiltonian with the bulk magnetization mass Eq.\eqref{Weyl}.       

\begin{figure}
\includegraphics[width=1\columnwidth]{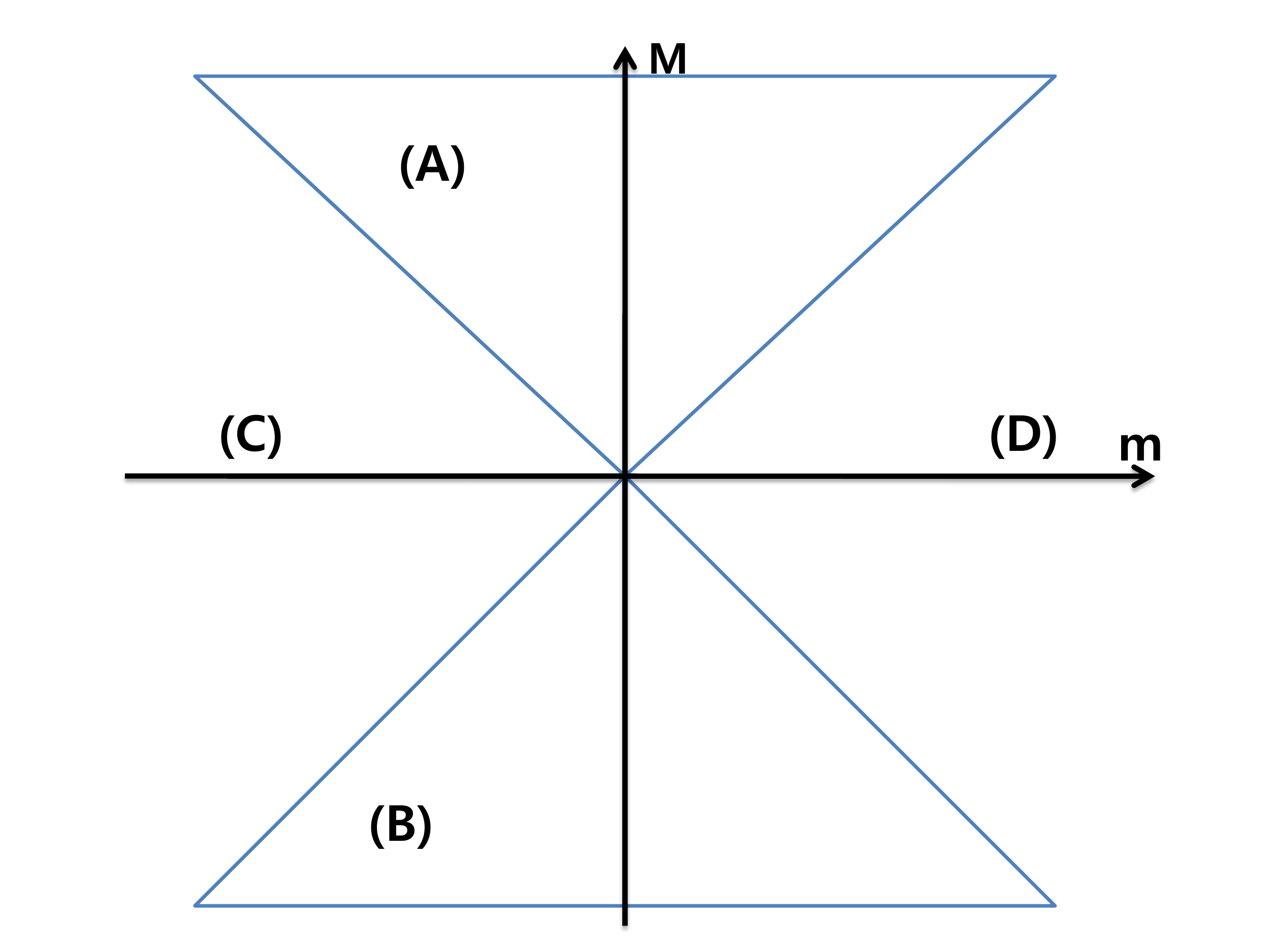}
\caption{Proposed phase diagram in terms of $(m, M)$ for $b<0$ in Hamiltonian Eq.\eqref{Weyl}: (A),(B) Topological band gap $M$ is larger than magnetization mass $m$, thus we are in the insulating phase. We have a topological band insulator (A) for $M>0$ and trivial insulator (B) for $M<0$. (C),(D) Magnetization mass is stronger than the topological band gap, and we have a Weyl semimetal phase. Note that if $m \rightarrow -m$, then two Weyl points change the sign of the chirality. At the transitions $|m|=|M|$, two Weyl points meet each other and result topological or trivial insulators.}
\label{Fig1}
\end{figure} 

- {\it Energy spectrum of the band theory}: The general behavior of the spectrum of the Hamiltonian Eq. \eqref{Weyl} can be easily read off from the following commutation relations $[\sigma^{z}\tau^{0}, \tau^{x}] =0$ and $[\tau^{x}, \tau^{z}\sigma_{\mu}]\neq 0$. Due to the commutation $[\sigma^{z}\tau^{0}, \tau^{x}] =0$, we will have the competition between the magnetization mass $m$ and the topological mass $M_{0}$. Explicitly, we have the energy spectrum $E({\bf k})$ for general $m$ and $M({\bf k})$
\beq
E ({\bf k}) = \pm \sqrt{v^{2}(k_{x}{}^{2}+k_{y}{}^{2}) + (m \pm \sqrt{M^{2}+v^{2}k_{z}{}^{2}})^{2}}
\label{Spec}
\eeq
we find that there are two mass gaps $\Delta_{\pm}$ where $\Delta_{+} = m + \sqrt{M^{2}+v^{2}k_{z}{}^{2}}$ and $\Delta_{-}= m - \sqrt{M^{2}+v^{2}k_{z}{}^{2}}$. We immediately see that the gap $\Delta_{-}$ can close if $m>M_{0}$ at the momentum $vk_{z} = \pm \sqrt{m^{2}-M(\bf k){}^{2}}$ which are the positions of Weyl points in the momentum space. We now discuss two cases where $m< M_{0}$ and $m>M_{0}$ (See Fig1).

- {\it Topological band and trivial insulator phases}: Here we discuss the case $m<M_{0}$. To study these phases, we use the adiabatic argument, i.e., we are in the same topological class as far as the gap remains open at ${\bf k}=0$ (we assume that ${\bf k}=0$ point is the only point which possibly closes the band gap) and the inversion symmetry is protected. As $\Delta_{+}$ is always positive for any $m$, our main focus goes to $\Delta_{-}$. The Dirac mass $\Delta_{-}({\bf k}=0) = m - M_{0}$ which closes at $m = M_{0}$, signaling quantum phase transition to the Weyl semimetal phase. When $m<M_{0}$, we need to further specify the mass $M({\bf k}) = M_{0} + b {\bf k}^{2}$ and have a topological band insulator phase due to the band inversion if $M_{0}b<0$ and a trivial insulator phase if $M_{0}b>0$. When we are in the topological band insulator phase (and also the chemical potential lies in the surface magnetization gap $ \sim m<M_{0}$), we have Chern-Simon effective surface theory with external electromagnetic gauge $A_{\mu}$  
\beq
L_{eff} = \frac{|m|}{2m}\times\frac{1}{4\pi} A_{\mu}\partial_{\nu}A_{\lambda} \varepsilon^{\mu\nu\lambda}
\eeq   
which is the quantum anomalous Hall effect with the {\it quantized} hall coefficient~\cite{fkm, axion_Essin, qi1} $\sigma_{xy} = \pm  e^{2}/2h$ . 

- {\it Gapless Weyl semimetal phase}: We now proceed to the case $m>M_{0}$ where the mass $\Delta_{-}({\bf k}=0)<0$. For the low-energy physics, we ignore the quadratic dependence on $M({\bf k}) = M_{0} + O({\bf k}^{2})$ to clarify the physics of this phase. Then, we have two Dirac nodes in $O({\bf k})$ where $vk_{z} = vK_{c,\pm} = \pm vK_{c} = \pm \sqrt{m^{2}-M_{0}{}^{2}}$. Near the critical points $k_{z}\sim \pm K_{c}$, the low-energy theory of the Hamiltonian Eq.\eqref{Weyl} reduces into (with $q_{z,\pm} = k_{z}-  K_{c,\pm}$)
\beq
{\bf H (\bf k)} = \epsilon ({\bf K_{c}} ) + 
\left[ 
\begin{array}{cc} 
u_{\pm}q_{z,\pm}& vk_{-}  \\
vk_{+} & -u_{\pm}q_{z,\pm}   
\end{array} 
\right]
\label{Weyl2}
\eeq
with anisotropic Dirac speed along ${\hat z}$ direction, $u_{\pm} = \mp v^{2}|K_{c}|/m$. This Hamiltonian is nothing but the Weyl fermion with the winding number $\pm 1$ localized at $K_{c,\pm}$ (two Weyl points carry the chirality $c_{\pm} =$ sign$(u_{\pm}) = \mp 1$ at the Dirac point~\cite{Ashvin}).

Due to this non-zero winding number, the Weyl Hamiltonian Eq.\eqref{Weyl2} has an anomalous Hall effect~\cite{Burkov, Yang, Fang} in $xy$- plane. Because each layer of the whole {\it bulk} participates in anomalous Hall effect, we define the anomalous Hall response {\it per layer}. The simple computation shows that the first chern number $C_{k_{z}} = \pm 1$ for $k_{z} \in (-K_{c}, K_{c})$ with $K_{c} = |K_{c,\pm}|$. Then, this would support the anomalous Hall conductance (when the chemical potential is exactly at the Dirac point)
\beq
\sigma_{xy} = \frac{e^{2}}{2\pi h} 2K_{c}
\eeq
This effect can be defined up to modular $e^{2}/h$ due to the ambiguity~\cite{Burkov, Yang} in $|K_{c}|$ which is up to modular $\pi$ (half of the size of the reciprocal vector along ${\hat z}$).

When two Weyl points are brought to the Brillouin zone center and pair-annhilate~\cite{Ashvin} ($m \rightarrow M_{0}$), we have topological insulator phase (characterized by axion angle $\theta = \pm \pi$) or trivial insulator phase ($\theta= 0$). The two axionic response of the resulting insulator phases can be anticipated from the inversion symmetry encoded in Eq. \eqref{TI} and Eq. \eqref{Weyl} because the angle $\theta \rightarrow -\theta$ under the inversion. By noting the following two reasonings ($1$) the axion angle $\theta = \pm \pi$ ($\theta = 0$) for topological insulator phase (trivial insulator phase) at $m\rightarrow 0$ and ($2$) the angle is tied to topological structure of the band structure ~\cite{parity1,parity2} and thus cannot be changed smoothly as far as the gap remains open, we conclude that the full region of topological band insulator (trivial insulator) phase in the phase diagram is characterized by $\theta = \pm \pi$ ($\theta = 0$). If the inversion symmetry is broken down, then the axion angle $\theta$ is no longer protected and expected to be continuously varying in terms of the parameters of the system.

It's also interesting to note that we can realize quantized layered anomalous Hall effects if the two Weyl points are dragged ($K_{c} = \pi$) to the Brillouin zone bounday~\cite{Burkov} and pair-annhilated. Then each layer in the bulk carries the Hall conductance 
\beq
\sigma_{xy} = \frac{e^{2}}{h}
\eeq 
and each layer forms integer quantum Hall states without magnetic field. From this consideration, we could remarkably obtain various interesting known topological phases, topological band insulator (gapped), Weyl semimetal phase (critical), and three-dimensional anomalous Hall (gapped) phase, within the single effective theory of the topological band insulator. In fact, we can go further if we notice that the same Hamiltonian can describe the two-dimensional topological band insulator. When the sample width of the three-dimensional topological insulator is reduced, then the material turns into the two-dimensional topological insulator~\cite{DimCross1} (with some oscillatory behavior as the function width). Furthermore, it's known that the two-dimensional topological insulator materials can be turned into the anomalous Hall effect upon doping of the magnetic materials~\cite{QAH, QAH1, axion_monopole, Franz_FM} Hence, it's plausible to connect the three-dimensional Weyl semimetal phase to the two-dimensional anomalous Hall effect in the sense of the dimesional cross-over~\cite{DimCross1}.  

From the above general discussion, we raise an interesting question: what will be the fate of the Weyl phase under reducing the sample width? To answer the question carefully, we consider that the sample width is reduced along ${\hat z}$, i.e., $L_{z}$ varies from some macroscopically large value (three-dimensional bulk limit) to a few layers (two-dimensional limit) and there are two cases we'd like to study: A. magnetization is orthogonal to the plane of the thin film (${\vec m} = m {\hat z}$), B. magnetization is in the plane of the sample (${\vec m} = m {\hat x}$). We assume that the magnetization strength $m$ is fixed while the sample width reduces. In the three-dimensional limit, we know that the both cases generate the Weyl semimetal phase if the magnetization $m$ is large enough. In the following discussion, we will find two different thin-film limits depending on the direction of ${\vec m}$. When the sample width $L_{z}$ is reduced enough, then $k_{z}$ starts to be discretized and each band can be labelled by $n_{z} \in {\bf Z}$ i.e., the electrons form the standing wave $\sim \sin(n_{z}\pi z/L_{z})$ in ${\hat z}$. As the lowest energy band $n_{z} =1$ is enough to describe the physics for the thin film limit, we can simply replace $k_{z}$ by $\pi/L_{z}$ in the effective theory Eq.\eqref{Weyl}. We begin with the case B (${\vec m} = m {\hat x}$) and the theory for the thin film is 
\beq
H = v \tau^{z} \sum_{a=1,2} \sigma_{a} \cdot {\bf k^{a}} + M({\bf k}) \tau^{x} + m\sigma^{x}\tau^{0}
\eeq
which has two Dirac points splitted in the momentum space along $k_{x}$. Specifically, two Dirac points are at $k_{x,c} = \pm \frac{\sqrt{m^{2}-M^{2}}}{v}$ and we have the gapless phase. If the magnetization mass is not strong enough, we are in the trivial insulator phase as expected. For the case A (${\vec m} = m {\hat z}$), we would find the theory
\beq
H = v \tau^{z} \sum_{a=1,2} \sigma_{a} \cdot {\bf k^{a}} + M({\bf k}) \tau^{x} + m\sigma^{z}\tau^{0}
\eeq
which is {\it gapped} everywhere. In fact, this is the model for the anomalous Hall effect if $|m| > |M|$ (otherwise, it's trivial), and the model supports the quantized anomalous Hall response $\sigma_{xy} = \frac{e^{2}}{h}$. So, we conclude that the Weyl semimetal phase for the case A (${\vec m} = m {\hat x}$) turns into the anomalous Hall effect if $|m|>|M({\bf k})|$ (or trivial insulator) as the sample width reduces. Note that while the sample width reduces, the bulk gap $M({\bf k})$ oscillates~\cite{DimCross1} as the function of the width $L_{z}$ and thus we expect to have the oscillation between anomalous Hall effect and trivial insulator. 

- {\it Other commuting masses and possible gapless phases}: The realization of Weyl semimetal phase in this paper rely on the commutation relation between mass/kinetic terms in Hamiltonian Eq.\eqref{Weyl}. In general, if two Dirac masses $Q$ and $M$ satisfy $[Q,M] =0$, then two masses tend to {\it kill} each other. Given the topological mass $M \sim \tau^{x}$ and the kinetic term $K\sim \tau^{z}\sigma^{a}$ in Eq. \eqref{Weyl}, we can classify the mass terms $Q$ that gives the possible gapless phases from the topological band insulator. The conditions are $[Q, M] = 0$ and $[K_{i}, Q] \neq 0$ for at least one of $i=1,2,3$. Then these conditions include $Q= ({\hat m} \cdot {\hat \sigma})\tau^{0}$ which generates `two' Weyl points, and $Q = \tau^{x}\sigma_{i}$ which generates a gapless cylinder around the origin, e.g., $ k_{x}{}^{2} + k_{y}{}^{2} = k_{c}{}^{2}$ along $k_{z}$ for $Q= \tau^{x}\sigma^{z}$. However, the second case would be easily gapped out in perturbation. So the magnetization mass is the only mass that is capable of inducing the Weyl semimetal phase in competion with the topological band gap.

\section{On the real material}
We begin with the basic model for Bi$_{x}$Sb$_{1-x}$ with the magnetic mass due to the Zeeman coupling $\sim J{\bf m} \cdot {\bf S}$ where ${\bf m}$ is the polarized magnetic moment in the system. Even though ${\bf m}$ can polarize in any direction, we further assume ${\bf m}$ to be parallel with ${\hat z}$. This can be prepared by applying the external magnetic field in ${\hat z}$ when the magnetic order sets into the sample. So, we assume $m_{x,y} << m_{z}=m$ and further ignore $m_{x,y}$. We begin with the standard tight-binding model on the diamond lattice~\cite{fkm} for the topological band insulator 
\beq
H_{TI} = t \sum_{<i,j>} c^{\dagger}_{i}c_{j} + i\lambda_{SO} \sum_{<<i,j>>} c^{\dagger}_{i} {\bf s} \cdot {\hat e}_{ij} c_{j}
\label{Dia}
\eeq
where the second term encodes the spin-orbit coupling which connects the next nearest neighbors. The Hamiltonian has the inversion symmetry and time reversal symmetry. The spectrum of the Hamlitonian Eq.\eqref{Dia} is very well understood and is topological band insulator (with the distortion). Now, we introduce the magnetization over the sample and assume that the doped magnetic material approximately induces the uniform on-site Zeeman term  
\beq
H_{z} = m \sum_{i} c^{\dagger}_{i}{\bf s}^{z} c_{i}
\eeq
We move to the momentum space for the Hamiltonian $H_{TI}+H_{z}$ and find 
\beq
H = \sum_{i=1}^{5} d_{i}({\bf k}) \Gamma^{i} + m \Gamma^{34}
\eeq
where we followed the notation of Liang Fu {\it et.al.}~\cite{fkm} We can diagnolize it to obtain the band structure 
\beq
E({\bf k}) = \pm \sqrt{d_{3}^{2} + d_{4}^{2} + (m \pm \sqrt{d_{1}^{2} + d_{2}^{2} + d_{5}^{2}})^{2}}
\label{Sp}
\eeq
Note the remarkable resemblance of the above energy spectrum Eq.\eqref{Sp} to the band spectrum Eq.\eqref{Spec} which we obtained from the continuum model. As $H_{z}$ is independent of the momentum ${\bf k}$, we don't need to solve for all momentum ${\bf k}$ but we can simply look at the points where the gap for $H_{TI}$ Eq.\eqref{Dia} is small and where $d_{3}({\bf k}), d_{4}({\bf k})$ and $d_{5}({\bf k})$ go as $\sim |{\bf k}| + O({\bf k}^{2})$. There are eight such points ($\Gamma$ point, three $X$ points, and four $L$ points) in the Brillouin zone and we will have Weyl points out of these eight points. For example, we find that there are two Weyl points near the $\Gamma$ point if $m > m_{c} = \sqrt{m^{2}-(4t)^{2}}$ at 
\beq
{\bf k}_{c} = (0,0, \pm \sqrt{\frac{m^{2}-(4t)^{2}}{t^{2}}})
\eeq
where $|4t|$ is the band gap at the $\Gamma$ point. We find that this simple tight-binding model agrees with the continuum calculation, and hence we've provided that the microscopic lattice model for the Weyl phase.  

- {\it experimental preparation}: In the experiment, topological band gap $M$ tends to be the largest energy scale in the system ($\sim 0.3$eV for Bi$_{2}$Se$_{3}$~\cite{BiSe}) and this would act as the drawback to realize the Weyl semimetal phases in that the magnetization mass $m$ should win against the band gap $M$. Hence it's desirable to reduce the band gap by bringing the system to near the critical point ($M \rightarrow 0$) between topological and trivial insulators. Recently, the mass gap of the topological insulator material TlBi(S$_{1-x}$Se$_{x}$)$_{2}$ appears to be under the control and could be tuned to the criticality in experiment~\cite{BiSeS}. Though it's difficult to predict the energy scale for the magnetic mass, some theoretical calculations ($\sim 5$ - $50$meV) and experiments ($\sim 0.1$eV) seems to be positive~\cite{QAH, QAH1} to see this physics (at least in the thin film limits). Due to the disorders, it might not be easy to detect the transition itself. However, if the magnetic mass is large enough, Weyl phase should appear. If there are more than one Dirac point in the low energy, there could be more than two Weyl points (depending on the precise form of the `representation' of the magnetization mass at the Dirac point).   

In terms of observation, the best way to detect Weyl semimetal phase would be perhaps ARPES experiment to see the Weyl points and the strange surface states. Transport experiment would be also applicable to detect anomalous Hall effects. Note that $\sigma_{xy} \sim k_{c} \sim (m^{2}-M^{2})^{1/2}$ where $M$ is tunable by controlling the width of the quantum well. Via this, $\sigma_{xy}$ would be tunable in the thin film limit. For the reasobale prediction with the magnetization gap $\sim 50$meV in thin film limit and the optimal preparation of the sample $M_{0} \rightarrow 0$, we predict $\sigma_{xy} \sim O(10^{1})$  $\Omega^{-1}$cm${}^{-1}$. 

{\it Note:} Upon completion of this work, we became aware of a paper by A.A. Burkov {\it et.al.}~\cite{Burkov2} where the realization of Weyl semimetal phase by magnetic perturbation in the topological insulator (at criticality between topological insulator and normal insulator phases) is also considered.  

\acknowledgements 
G.~Y.~C thanks Eun Gook Moon, Meng Cheng, Roger S.K. Mong, and Su Yang Xu for helpful comments. G.~Y.~C. also thanks Joel E. Moore for support and encouragement during this work and acknowledges support from NSF DMR-0804413 and a KITP Graduate Fellowship.

\bibliography{Weyl_jem}
\end{document}